\documentclass[conference, a4paper]{IEEEtran}
\IEEEoverridecommandlockouts
\usepackage{cite}
\usepackage[export]{adjustbox}
\usepackage{amsthm,amsmath,amssymb,amsfonts}
\usepackage{algorithmic}
\usepackage{graphicx}
\usepackage{textcomp}
\usepackage{xcolor}
\usepackage{float}
\usepackage{subfigure}
\usepackage{subcaption}

\usepackage[ruled,vlined]{algorithm2e}
\usepackage{bbold}
\usepackage{array}
\usepackage{booktabs}
\usepackage{mathrsfs}
\usepackage[normalem]{ulem}

\usepackage{todonotes}
\usepackage{url}

\def\BibTeX{{\rm B\kern-.05em{\sc i\kern-.025em b}\kern-.08em
    T\kern-.1667em\lower.7ex\hbox{E}\kern-.125emX}}

\begin{document}

\title{Probabilistic Mobility Load Balancing for Multi-band 5G and Beyond Networks\\
{}
}

\author{
    \IEEEauthorblockN{Saria Al Lahham, Di Wu, Ekram Hossain, Xue Liu, Gregory Dudek}
    \IEEEauthorblockA{Samsung AI Center, Montréal, Canada
    \\\{s.allahham, di.wu1, steve.liu, greg.dudek\}@samsung.com, \{e.hossain\}@partner.samsung.com}
    }


\maketitle

\begin{abstract}
The ever-increasing demand for data services and the proliferation of user equipment (UE) have resulted in a significant rise in the volume of mobile traffic. Moreover, in multi-band networks, non-uniform traffic distribution among different operational bands can lead to congestion, which can adversely impact the user's quality of experience. Load balancing is a critical aspect of network optimization, where it ensures that the traffic is evenly distributed among different bands, avoiding congestion and ensuring better user experience. Traditional load balancing approaches rely only on the band channel quality as a load indicator and to move UEs between bands, which disregards the UE's demands and the band resource, and hence, leading to a suboptimal balancing and utilization of resources. To address this challenge, we propose an event-based algorithm, in which we model the load balancing problem as a multi-objective stochastic optimization, and assign UEs to bands in a probabilistic manner. The goal is to evenly distribute traffic across available bands according to their resources, while maintaining minimal number of inter-frequency handovers to avoid the signaling overhead and the interruption time. Simulation results show that the proposed algorithm enhances the network's performance and outperforms traditional load balancing approaches in terms of throughput and interruption time.

\end{abstract}


\section{Introduction}
Mobility load balancing (MLB) has been recognized as a crucial element of managing radio resources \cite{tolli2002performance}, and a core problem in 5G communication networks. In fact, MLB is commonly studied within the context of self-organizing networks (SON), which provide a comprehensive framework for self-optimization and self-healing, with load balancing being a critical aspect of the self-optimization component \cite{3gpp.36.902}. MLB in 3GPP standards is defined as moving cell-edge user equipment units (UEs) in a crowded cell to a less-crowded neighboring cell. In multi-band networks, MLB can additionally move UEs between the available bands via inter-frequency handovers (HOs) based on the channel quality for each band, where it aims to enhance the UEs throughput, ensure fairness between UEs, and minimize the latency \cite{hu2010self}. Given the continuously increasing number of UEs and their increased data demand, coupled with their demand heterogeneity and mobility, load balancing is becoming even more important in managing network resources. Moreover, it is reported that mobile network data was doubled in the last two years alone, and with the introduction of extended-reality (XR) applications and advancements in the quality and resolution of video content, it is expected that network traffic will increase by at least 38\% over the next years \cite{ericsson_2022}. This shows the need for efficient load balancing techniques that can effectively distribute traffic across available resources to ensure optimal network performance.

The existing literature on MLB focuses on the optimal adjustment of the HO controllable parameters, specifically, the cell individual offsets (CIOs), relying on the current load of each cell and the measurements reported by the UEs such as the Reference Signal Received Quality (RSRQ). An example of the usage of CIOs is the A3 event as defined in the 3GPP standards \cite{3gpp.36.331}:
\begin{equation}
\text{A3}:  R_n - R_s > O_{n,s} + H,
\end{equation}
where $R_n$ and $R_s$ are the RSRQ reported values for a neighboring cell and the serving cell, respectively, $H$ is the hysteresis, and $O_{n,s}$ is a controllable CIO. Various rule-based MLB techniques were proposed in the literature. For instance, the authors in \cite{kwan2010on} proposed an algorithm to adjust the CIO between cells based on the difference in their load, i.e., a cell which has lesser load than its neighbor its CIO gets decreased by a factor to ease the HO triggering from its neighbor, where the factor is a function of the difference between the cells' loads. A similar approach was proposed in \cite{yang2012high}, but the authors considered adding or subtracting
an adaptive step-size to the offset based on the time average of the load for each cell. 
Furthermore, to tackle the problem of non-uniform traffic
distribution between cells, the authors in \cite{huang2015threshold} proposed a traffic and cell load balancers with a new HO rule-set, including adjustable CIOs and time-to-trigger based on the average RB consumption of each UE.
Apart from rule-based methods, numerous learning-based methods were proposed and were shown to be superior to rule-based methods. For example, Reinforcement Learning (RL)-based MLB was presented in \cite{kang2021hierarchical} in addition to inactive mode load balancing. The authors proposed a hierarchical RL algorithm to control both MLB CIOs for active UEs, and the cell re-selection parameters for inactive UEs to mitigate overloading cells, which resulted a much better performance. To generalize on real-world data and to enable better tradeoffs between different key performance indicators (KPIs), the authors in \cite{feriani2022multiobjective} proposed a multi-objective meta-RL algorithm. Compared to rule-based methods and single objective RL, the proposed approach achieved better KPIs and Pareto fronts. Furthermore, a set of RL algorithms were proposed in \cite{alsuhli2021mobility} with a diverse set of reward functions to satisfy the operators’ needs and key performance
indicators (KPIs). Each algorithm and reward function target a different scenario based on the density of the cells and the UEs traffic patterns.

Nonetheless, CIO-based HO for MLB approaches do not move UEs between bands according to their demands or band resources with inter-frequency HOs, but they rely on the fact that the lesser the load of a band, the better the channel quality (e.g., RSRQ). However, once the band resources (i.e., Physical Resource Blocks (PRBs)) are fully utilized, the channel quality becomes independent of the current load, and performance of aforementioned approaches deteriorates. To this end, in this work, we consider a centralized approach by employing forced HOs, where a base station (BS) can change the operating band of a UE through an RRC reconfiguration message \cite{3gpp.38.300}. By doing so, the BS have the control over moving UEs to more appropriate bands based on the UE's traffic and the bands' available resources. As such, we formulate the load balancing problem as a multi-objective UE-band assignment problem.
To mitigate the signaling overhead and reduce the interruptions time caused by the forced HOs, we set our objective as the joint minimization of the number of inter-frequency HOs and the bands' total load. 
Moreover, to avoid the computational complexity overhead caused by solving the assignment problem, we propose an event-based algorithm that checks if there is a necessity for load balancing. The contributions of this work can be summarized as follows:
\begin{itemize}
    \item We formulate a multi-objective stochastic optimization problem to model the UE-band assignment problem, aiming at minimizing the maximum load of the bands and number of HOs required.
    \item By leveraging problem transformation methods (e.g., epigraph technique), we reduce the problem into a Linear Program to reduce the solution complexity.
    \item We propose an event-based MLB algorithm with probabilistic assignment scheme, which uses the solution of the Linear Program, such that the assignment for a UE is given by a probability distribution over the bands.
\end{itemize}

The remainder of the paper is organized as follows: Section II
introduces the considered system model in this work. Section III presents the formulated multi-objective optimization problem, while Section IV presents the solution methodology and the proposed algorithm. In Section V we show the analysis for the solution and the
performance evaluation. Finally, Section VI concludes the paper.

\section{System Model}\label{sys_model}

We consider a cellular network consisting of $N$ macro-gNodeBs (gNBs), each of which covers three different cells. Each cell operates over $B$ non-overlapping frequency bands, and serves $U$ UEs that are uniformly distributed across the cell. Within a cell, the set of available bands is denoted by $\mathcal{B}$ and the set of UEs is denoted by $\mathcal{U}$. Throughout this work, we omit the cell index since we focus on the cell-level multi-band MLB, and each cell has its own unique set of UEs and bands. As for MLB between different cells, the 3GPP default rule set for MLB \cite{3gpp.36.902} is employed such as the A2 and A5 events.
The downlink (DL) channels for each band to each user in a cell have time-varying capacities. A UE $u$'s DL channel rate from a band $b$ at time $t$ is random variable and denoted by $r_{u,b}(t)$.
We assume that $r_{u,b}(t)$ is chosen from a discrete set of rates as in a practical wireless system and sampled from a stationary distribution. Note that different UEs for each band can possibly have heterogeneous rate distributions, depending mainly on the band resources, denoted by $N_b$, and the Signal-to-Interference-Noise-Ratio (SINR), where band resources are usually represented by the number PRBs in that band. We then denote the $U\times B$ rate matrix in a cell as $\mathbf{R}(t)$, such that the $u^{th}$ row $\mathbf{r}_u(t)$ denotes the rate vector for a UE $u$ across all bands, and the $b^{th}$ column $\mathbf{r}_b(t)$ denotes the rate vector in a band $b$ for all the UEs. Afterwards, we assume a full buffer traffic model, and define the UEs' demand vector as $\mathbf{d}(t)$, where the UE $u$'s demand at time $t$ is given by:
\begin{equation}
    d_u(t) =  a_u(t) s,
\end{equation}
in which $a_u(t)$ is a Poisson random variable with a mean of $\lambda_u$ and denotes the number of packets arrivals, and $s$ represents the packet size in bits. Thereafter, we define the load vector for a band $b$ as a vector as follows:
\begin{equation}
    \boldsymbol{\rho}_b (t) = \mathbf{d}(t) \oslash \mathbf{r}_b(t),
\end{equation}
where $\oslash$ is the Hadamard division operator (i.e., element-wise division). The load vector defines how much each UE increases the band load based on the UE's incoming traffic, channel quality, and DL rate from that band. Indeed, UEs with bad channel quality with a band means low DL rate, and hence, a higher load to the band (e.g., occupying more PRBs to meet the UE demand). For the remainder of the paper, the time index $t$ will be omitted for the sake of clarity.

\section{Problem Formulation}\label{prob_formulation}
Our main objective in this work is to balance the loads between the available bands by moving UEs between bands via forced inter-frequency HOs while minimizing the number of needed HOs. As such, we first define the $U \times B$ UE-band assignments matrix $\mathbf{X}$, such that the $u^{th}$ row $\mathbf{x}_u$ denotes the assignment vector for a UE $u$ between all bands, and the $b^{th}$ column $\mathbf{x}_b$ denotes the assignment vector in a band $b$ for all the UEs. The assignment entries $x_{u,b} = 1$ if the UE $u$ is assigned to a band $b$, and $x_{u,b} = 0$ otherwise. Subsequently, we define the load balancing objective function as follows:
\begin{equation}
	f_1(\mathbf{X}) = \max({\mathbf{x}^\top_b \mathrm{E}\left[ \boldsymbol{\rho}_b \right ]: b \in \mathcal{B} }),
\end{equation}
which represents the maximum load across all the bands, and $\mathrm{E}\left[ \boldsymbol{\rho}_b \right ]$ denotes the expected load vector for a band. In fact, the optimal solution for minimizing the maximum load is to keep all the loads the same for all the bands, which is our target in MLB. Afterwards, we define the number of inter-frequency HOs objective function as follows:
\begin{equation}
	 f_2(\mathbf{X}) = \left\| \mathbf{X} - \hat{\mathbf{X}} \right\|_1,
\end{equation}
where $||.||_1$ for a matrix here denotes the absolute sum of all the entries, and $\hat{\mathbf{X}}$ is the UE-band assignments at $t-1$ prior to the load balancing. It is important to notice that there is a trade-off between the two objectives. On the one hand, the optimal value for the second objective is achieved when 
$\mathbf{X} = \hat{\mathbf{X}}$, indicating that no UE needs to be relocated to another band, thereby precluding any load balancing with no HOs. On the other hand, in a worst-case scenario, it may be essential to relocate all UEs from their current band to another one, resulting in a much worse solution for the second objective. To this end, we formulate a scalarized weighted multi-objective optimization problem (MOP) as follows:
\begin{subequations}\label{eq:P1}
\begin{alignat}{2}
&\!\min_{\mathbf{X}}        &\qquad& w f_1(\mathbf{X}) + (1-w)f_2(\mathbf{X}) \label{eq:P1obj}\\
&\text{s.t.} &      &  \left\|  \mathbf{x}_u\right\|_1 = 1,\quad \forall u \in \mathcal{U}\label{eq:P1constraint1}\\
&                  &      &  \mathbf{x}^\top_u  \mathbf{r}_u \geq r_{\min},\quad \forall u \in \mathcal{U}\label{eq:P1constraint2}\\
&                  &      & \left\|  \mathbf{x}_b\right\|_1 \leq n_{\max},\quad \forall b \in \mathcal{B} \label{eq:P1constraint3}\\
&                  &      &\mathbf{X} \in \{ 0,1 \}^{B\times U}, \label{eq:P1constraint4}
\end{alignat}
\end{subequations}
where $w \in[0,1]$ is the objective weight hyper-parameter in the MOP, constraint (\ref{eq:P1constraint1}) tells that a UE can only be assigned to one band, constraint (\ref{eq:P1constraint2}) ensures that the UE will be granted a minimum rate of $r_{\min}$, constraint (\ref{eq:P1constraint3}) avoids over-congesting bands that has better channel quality and offers higher data rates by setting a cap $n_{\max}$ on the number of UEs on this band, such that $n_{\max}$ is proportional to the bands' resources, and finally, constraint (\ref{eq:P1constraint4}) tells that the decision variable is a binary matrix.
It can be noticed that the problem is not only an NP-Hard Integer Nonlinear Program \cite{boyd2004convex}, but also a stochastic optimization problem by having the UEs' data rate and the incoming traffic as random vectors in $f_1(\mathbf{X})$. To tackle these challenges, common approaches such as relaxing the integer decision variables and estimating the random variables can be followed and further optimized as will be demonstrated in the next section.
\section{Methodology and Proposed Solution}\label{method}
This section presents the proposed approach for solving the formulated MOP that aims to redistribute UEs across bands, such that the maximum bands' load is minimized while reducing the number of needed inter-frequency HOs. 
\subsection{Problem Transformation}
First to address, in practice, a possible infeasibility can result from constraint (\ref{eq:P1constraint2}), where a UE can have data rates less than $r_{\min}$ due to bad channel conditions, and it does not satisfy the MLB rule set to move to a neighboring cell. As such, the best possible solution is to associate with the band $b$ that has the best channel quality as follows:
\begin{equation}
b = \arg \max_b ~\boldsymbol{\gamma}_{u},
\end{equation} where $\boldsymbol{\gamma}_{u} = \{\gamma_1,...,\gamma_B\}$ can be the set of a UE $u$ RSRQ values with each band. It is important to note that any relocated UE according the aforementioned solution adds a load to the associated band. As a result, we denote the incurred load to a band $b$ prior to the optimization as $\hat{\rho}_b$, then the UE is removed from the set of considered UEs $\mathcal{U}$ in the optimization.

Afterwards, to deal with random vectors in the problem, one common practice is to consider the expected value of the random variables from observed samples over a time period of $\Delta t$ as an estimation instead of instantaneous samples. Even though such estimation is not optimal, it works in practice in many scenarios \cite{birge2011introduction}. Accordingly, prior to solving the optimization, the load vector in $f_1(\mathbf{X})$ is substituted by $\boldsymbol{\rho}_b = \mathrm{E}\left[\boldsymbol{\rho}_b(t) \right ]$, where the expectation is considered over time. Nevertheless, the choice of $\Delta t$ is important, since having a relatively small or large values can lead to sub-optimal solutions. 

Subsequently, to address the integrality of the problem, we relax the integer variables to continuous ones in the range of $[0,1]$. We further propose the probabilistic rounding approach, which is to make use of constraint (\ref{eq:P1constraint1}) and regard the UE assignment vector $\mathbf{x}_u$ as a probability distribution over the bands. Not only this approach allows us to have better solutions and mitigate the deterministic rounding errors that might occur in the solution, but also takes into account the probabilistic nature of the problem. Accordingly, after solving the optimization, we assume for each UE there exist a probability distribution $\mathbf{p}_u$ that takes on the values of $\{1, 2, ..., B\}$, and has probabilities of $\mathbf{x}_u$, and then to select the assigned band $b$ we sample from that distribution, i.e., $b \sim \mathbf{p}_u$.

Last but not least, to transform the problem into an equivalent Linear Program (LP), we leverage the epigraph technique \cite{boyd2004convex} to linearize the objective function by first introducing a non-negative slack variable $t$, such that:
\begin{equation}
\mathbf{x}^\top_b \boldsymbol{\rho}_b \leq t,\quad \forall b \in \mathcal{B},
\end{equation}
and substituting $t$ instead of $f_1(\mathbf{X})$. Additionally, slack variables $y$ and $\mathbf{y}_b$ are introduced where:
\begin{subequations}\label{eq:P2}
\begin{alignat}{2}
&   &   &  y = \sum_{b \in \mathcal{B}} \left||\mathbf{y}_b |\right|_1,\\
&   &   &  \mathbf{x}_b - \hat{ \mathbf{x} }_b  \preceq \mathbf{y}_b,\\
&   &   &  - \left ( \mathbf{x}_b - \hat{ \mathbf{x} }_b  \right ) \preceq \mathbf{y}_b,
\end{alignat}
\end{subequations}
where $\preceq$ is the element-wise $\leq$ comparison. Consequently, we can formulate the equivalent transformed LP as follows:
\begin{subequations}\label{eq:P2}
\begin{alignat}{2}
&\!\min_{\mathbf{X}, \mathbf{y}_b, t, y}        &\qquad& w t + (1-w)y \label{eq:P2opt}\\
&\text{s.t.} &      &  \left\|  \mathbf{x}_u\right\|_1 = 1\label{eq:P2constraint1}\\
&                  &      &  \mathbf{x}^\top_u  \mathbf{r}_u \geq r_{\min},\quad \forall u \in \mathcal{U}\label{eq:P2constraint2}\\
&                  &      & \left\|  \mathbf{x}_b\right\|_1 \leq n_{\max},\quad \forall b \in \mathcal{B}\label{eq:P2constraint3}\\
&                  &      &\mathbf{x}^\top_b \boldsymbol{\rho}_b + \hat{\rho}_b \leq t,\quad \forall b \in \mathcal{B} \label{eq:P2constraint3}\\
&                  &      &  \mathbf{x}_b - \hat{ \mathbf{x} }_b  \preceq \mathbf{y}_b,\quad \forall b \in \mathcal{B}\label{eq:P2constraint4}\\
&                  &      &  - \left ( \mathbf{x}_b - \hat{ \mathbf{x} }_b  \right ) \preceq \mathbf{y}_b,\quad \forall b \in \mathcal{B} \label{eq:P2constrain5}\\
&                  &      &  \sum_{b \in \mathcal{B}} \left||\mathbf{y}_b |\right|_1 = y \label{eq:P2constrain6}\\
&   &      &  y \geq 0, t\geq 0, \mathbf{y}_b \succeq 0   ,\quad \mathbf{X}\in [0,1]^{B\times U}\label{eq:P2constrain7}
\end{alignat}
\end{subequations}

\begin{algorithm}[!ht]
	\caption{Probabilistic Mobility Load Balancing (PMLB)}\label{alg:two}
	\textbf{Input: $~\Delta t,~ L_{th},~ w,~ r_{\min},~ n_{\max}$}\;
	
	\For{$t=0:\infty$}{
	\ForEach{ $b \in \mathcal{B}$ }{
	 get $\mathbf{d}(t)$ and $\mathbf{r}_b(t)$\;
	 calculate $\boldsymbol{\rho}_b(t)$ according to (3)\;
	 store $( \mathbf{d}(t), \mathbf{r}_b(t), \boldsymbol{\rho}_b(t) )$\;
	}
	\eIf{ $t\mathbin{\%} \Delta t = 0$}{
			calculate LBI according to (11)\;
			\eIf{$\text{LBI} \geq L_{th}$ }{
			\textbf{continue}\;
			}{
			assign infeasible UEs according to (7)\;
			calculate $\mathrm{E}\left[\boldsymbol{\rho}_b(t) \right ]~ \forall b \in \mathcal{B}$\;
			$\mathbf{X} \gets$ solve the LP in (10) \;
			\ForEach{ $\mathbf{x}_u \in \mathbf{X}$ }{ sample a band $b$ with probabilities $\mathbf{x}_u$\;
			assign the band $b$ to UE $u$\;
			    }
			}
		}{\textbf{continue}\;}
		}
\end{algorithm}

In general, LP is a well-established and widely-used optimization model, for which many efficient and low-complexity algorithms have been developed, in addition to the existence of a variety of open sourced solvers such as CVXPY \cite{cvxpy} that can be used. It is important to mention that, in practice, the two objectives $t$ and $y$ are normalized in the range of $[0,1]$ by dividing over the maximum values of the objectives $f_1(\mathbf{X})$ and $f_2(\mathbf{X})$, respectively.

\subsection{Algorithm}
Two important concerns arise from our approach: Firstly, the optimal frequency at which the optimization should be performed and secondly, whether optimization is essential to be performed periodically in the first place. In fact, these two concerns are contingent upon numerous factors, but the most important ones are the mobility and data traffic patterns of the UEs. To address the concerns, we propose to perform the optimization every $\Delta t$ minutes, such that we satisfy an event based on the UE data traffic distribution across bands. Therein, to indicate how much the loads between the bands are evenly distributed, we define the load balancing index (LBI) as follows: 
\begin{equation}
    \text{LBI} = \frac{\left ( \sum_b \bar{\rho}_b \right)^2 }{B  \sum_b \bar{\rho}^2_b},
\end{equation}
where $\bar{\rho}_b= |\left| \boldsymbol{\rho}_b (t) |\right|_1 $ and represents the band $b$ load. One can notice that the LBI is simply the Jain's Fairness Index (JFI) \cite{jain1984quantitative}, but across the bands' loads. Usually, JFI is used to quantify how the resources are uniformly-distributed in resource allocation problems. In a similar manner, the JFI can also quantify how the loads are balanced since we aim to keep the bands' loads uniformly distributed. 
The LBI will vary eventually based on channel quality from each UE to their assigned band, and the movement of UEs when they enter and leave the cell since they either add or remove load from bands. Consequently, we define the event as the occurrence at which the LBI drops below a threshold, i.e., $\text{LBI} \leq L_{th}$. In fact, not only this event allows us to save some computation, but it also leads to less inter-frequency HOs since no UEs will be relocated if the event is not triggered. It is important to note that the optimization for each cell can be executed by the corresponding BS to perform load balancing between the bands in that cell. The proposed Probabilistic MLB (PMLB) algorithm is presented in Algorithm 1 and summarized as follows: For each band in a cell, get the measurement reports, the DL rate, and the incoming traffic size for each UE. Calculate the load that each UE is going to add to the band, and store the values. At each $\Delta t$, check the LBI event, if the loads are unbalanced, perform the feasibility checks and solve the LP in (10). Finally, assign UEs to a band based on the probability distributions that we obtained from solving the LP.

\addtolength{\topmargin}{0.03in}
\section{Experimental Results}\label{exp}
\begin{table}[t]
\caption{Simulation Parameters}
\centering
\normalsize
\begin{tabular}{ll}
\hline
Parameter               & Value                     \\ \hline
Simulation time          & 24 h                     \\
Inter-site distance      & 200 m                     \\
\# of cells             & 3                         \\
Number of bands $B$     & 4                         \\
Bandwidth for each band & {[}20, 10, 5, 10{]} MHz   \\
\# PRB for each band    & {[}100, 50, 25, 50{]} \\
Packet size             & 1500 Bytes                \\
Scheduler               & Proportional Fair         \\
$\Delta t$              & 2 min                     \\ 
LBI threshold $L_{th}$  & 0.8 \\ \hline
\end{tabular}\label{tab:sim_parameters}
\end{table}

The experiments are conducted on a proprietary System Level Simulator (SLS) \cite{feriani2022multiobjective}, which is a packet-level simulator that emulates traffic for 5G networks based on real-word data. The main simulation parameters are listed in Table \ref{tab:sim_parameters}. In the conducted experiments, we consider three dynamic traffic scenarios: \textbf{Scenario A:} A heavy traffic scenario with 400 UEs per cell and packet inter-arrival time $1/\lambda_u$ of 20 ms, \textbf{Scenario B:} A moderate traffic scenario with the same number of UEs but a packet inter-arrival time $1/\lambda_u$ of 50 ms, and \textbf{Scenario C:} A light traffic scenario with lower number of UEs with 200 UEs per cell and a packet inter-arrival time $1/\lambda_u$ of 50 ms. 
In what follows, we first analyze the proposed solution to the formulated MOP in terms of Pareto optimality and the proposed integer relaxation method. Afterwards, we compare the performance of the proposed algorithm against different baselines.
\subsection{Analysis of Optimization Solution}
One way to model MOP is by scalarization, where all the objectives are simply added but each objective is multiplied by a significance weight \cite{boyd2004convex}. To find the Pareto-optimal solutions to the MOP, the weights values are varied for each objective and then the solutions for each variation form the Pareto front. 
\begin{figure}[h!]
    \centering
\includegraphics[scale=0.6]{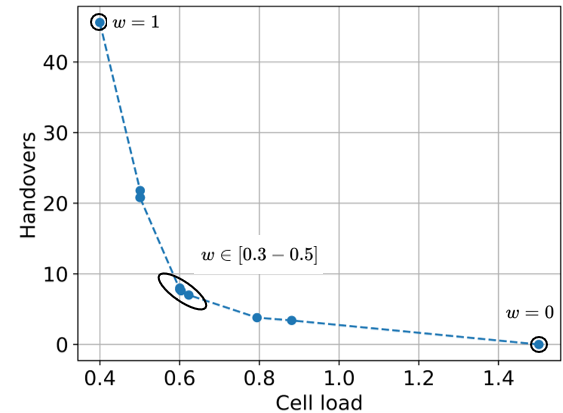}
    \caption{The Pareto Frontier of the MOP}
    \label{fig:pareto}
\end{figure}
The Pareto front for our MOP and the tradeoffs between the two objectives are depicted in Fig. \ref{fig:pareto}.
Each point on the curve represents a different value for the objective weight $w$. On the one extreme, when the weight for the load objective is equal to 0 (i.e., no significance and no load balancing needed), the HO objective achieves its optimal value, that is, no UEs are relocated. On the other extreme, when the weight for the load objective is equal to 1 (i.e., full significance), we achieve the lowest load possible, but the second objective achieves a very suboptimal value. It can be seen that the Pareto optimal solutions can be achieved when the weight $w$ is in the range of $[0.3,0.5]$. Then, we study the proposed integer relaxation approach in terms of time complexity (Fig. 2 (a)) and the objective value (Fig. 2 (b)). Such study helps to understand the reduction in time complexity and its decrease in the objective value due to the relaxation. To compare with our approach with probabilistic rounding (P-LB), first, the LP in (\ref{eq:P2}) is solved without integer relaxation using a Mixed-Integer LP (MILP) solver by employing the Branch and Bound algorithm. Afterwards, we solve it with integer relaxation considering the deterministic rounding approach (D-LP), where the values are rounded to the nearest integer (i.e., 0 or 1) and ties broken evenly. In Fig. 2(a), we show the LP solution time in milliseconds. It can be seen that as we increase the number UEs in the problem (i.e., increasing the number of variables), the time increases exponentially for MILP, whereas with the integer relaxations it increases only marginally. However, in Fig. 2(b), it can be seen that the MILP achieves the best objective value, and the D-LP method achieves a worse value, but it also increases the performance gap as we increase the number of UEs. Although for our approach we observe a slightly worse objective value, it keeps a constant gap between the MILP.

\subsection{Performance Comparison}
The proposed PMLB algorithm is compared to two baselines, namely, a rule-based MLB (RBMLB) and the 3GPP defined MLB, in addition to the no load balancing baseline. RBMLB is a proprietary MLB algorithm that is developed by Samsung and used by some telecom operators. It uses a fixed set of rules such that a pool of UEs that are randomly selected moved between bands, in addition to the A2, A4, and the A5 events to assess the channel quality prior to any HO. The NO MLB baseline let the UEs camp on the first 2 bands only without relocating any UEs, whereas the 3GPP defined MLB let the UEs associates with bands based on the A2 event. The performance comparison is shown in Fig. 3. All the results shown are the average metric per $\Delta t$ in a day, and averaged across all the cells.

\begin{figure}[htp]
\centering     
\captionsetup{justification=centering}
\subfigure[]{\label{fig:b}\includegraphics[scale=0.28]{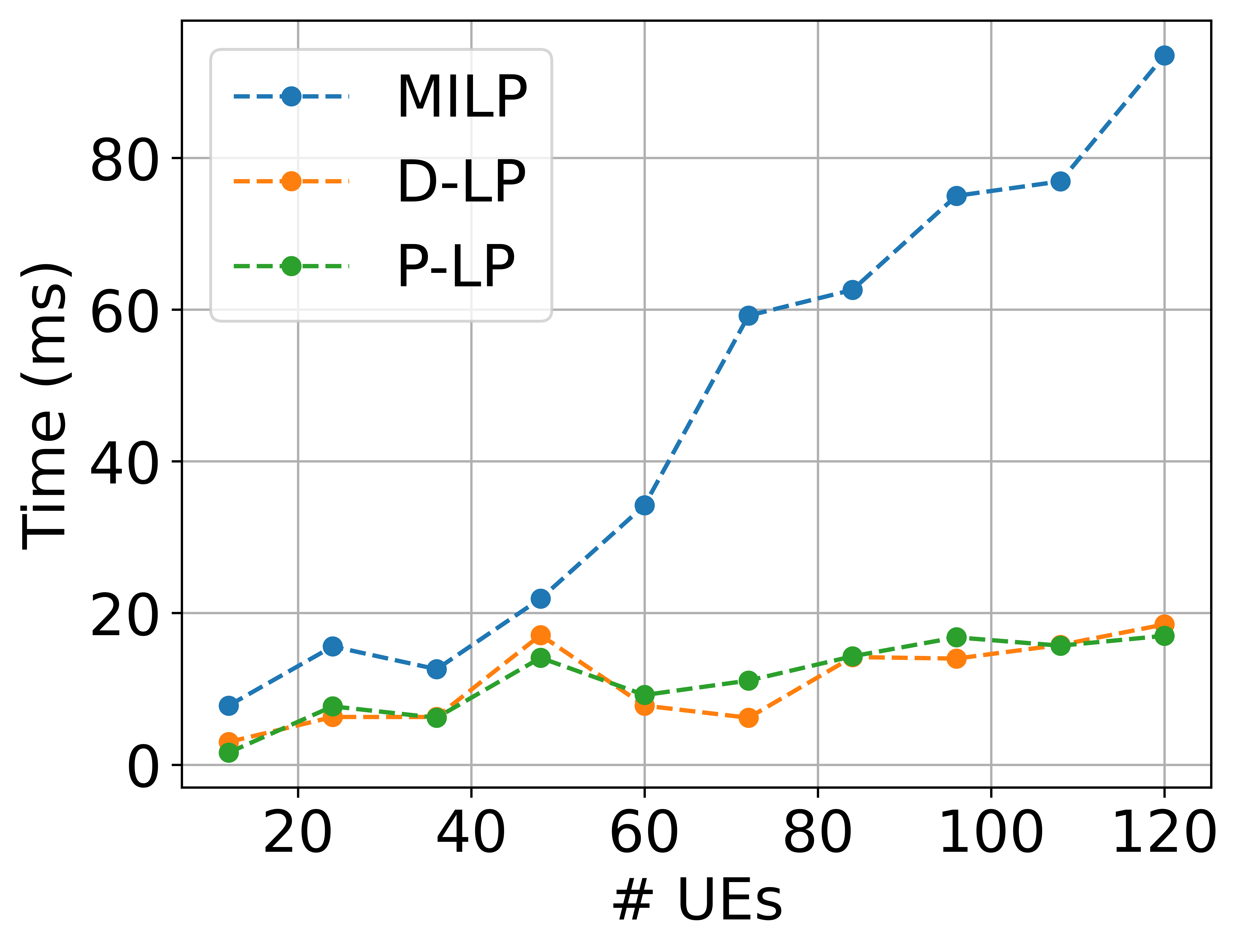}}
\subfigure[]{\label{fig:b}\includegraphics[scale=0.28]{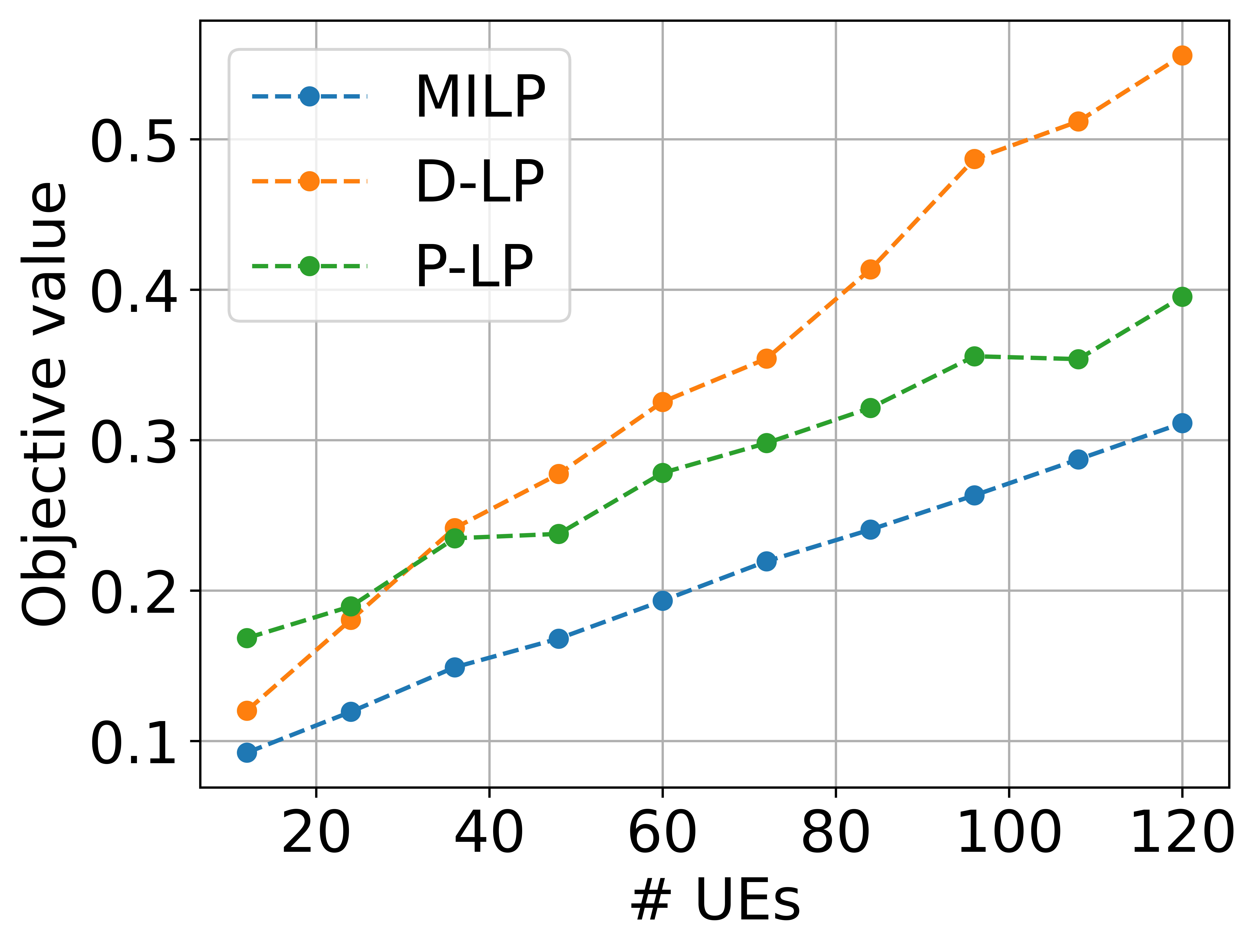}}
\caption{Optimization analysis in terms of (a) time complexity and (b) the objective function value between rounding techniques }
\end{figure}

\begin{figure*}[!ht]
\centering     
\captionsetup{justification=centering}
\subfigure[\hspace{-.3in}]{\label{fig:1}\includegraphics[scale=0.39,trim={0.2in 0 0.6in 0.54in},clip]{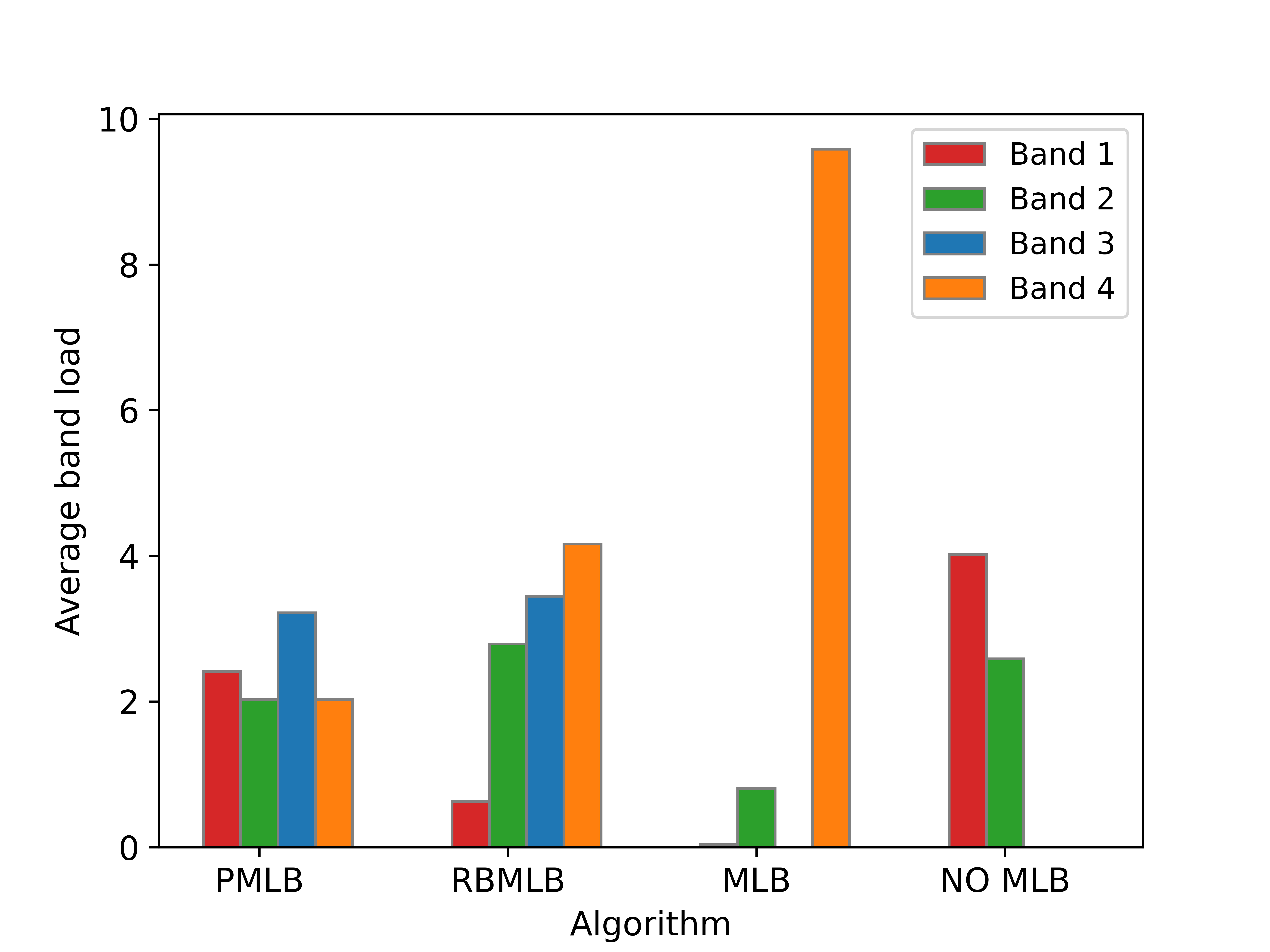}}
\subfigure[\hspace{-.3in}]{\label{fig:2}\includegraphics[scale=0.39,trim={0.2in 0 0.6in 0.54in},clip]{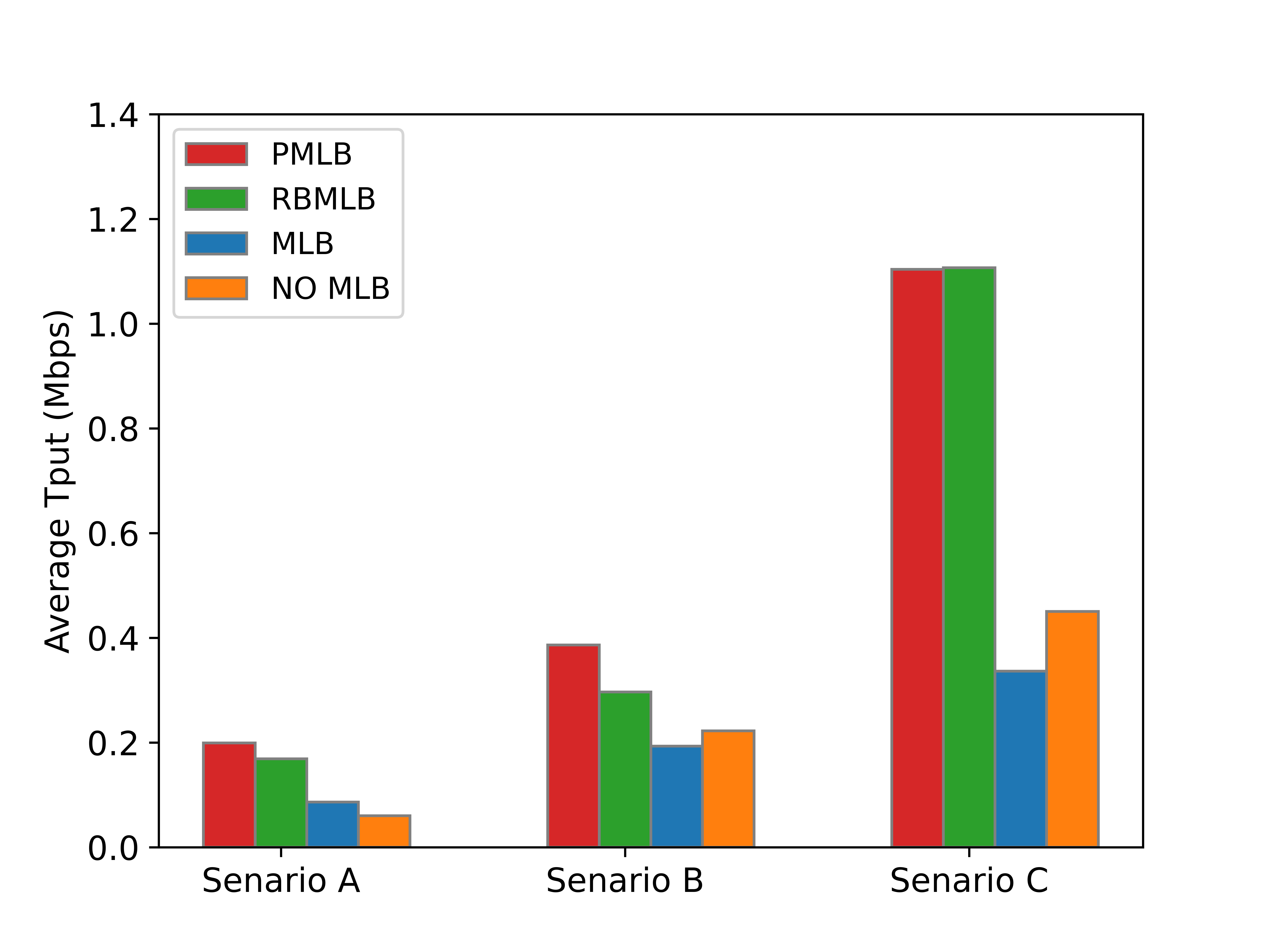}}
\subfigure[\hspace{-.3in}]{\label{fig:3}\includegraphics[scale=0.39,trim={0.2in 0 0.6in 0.54in},clip]{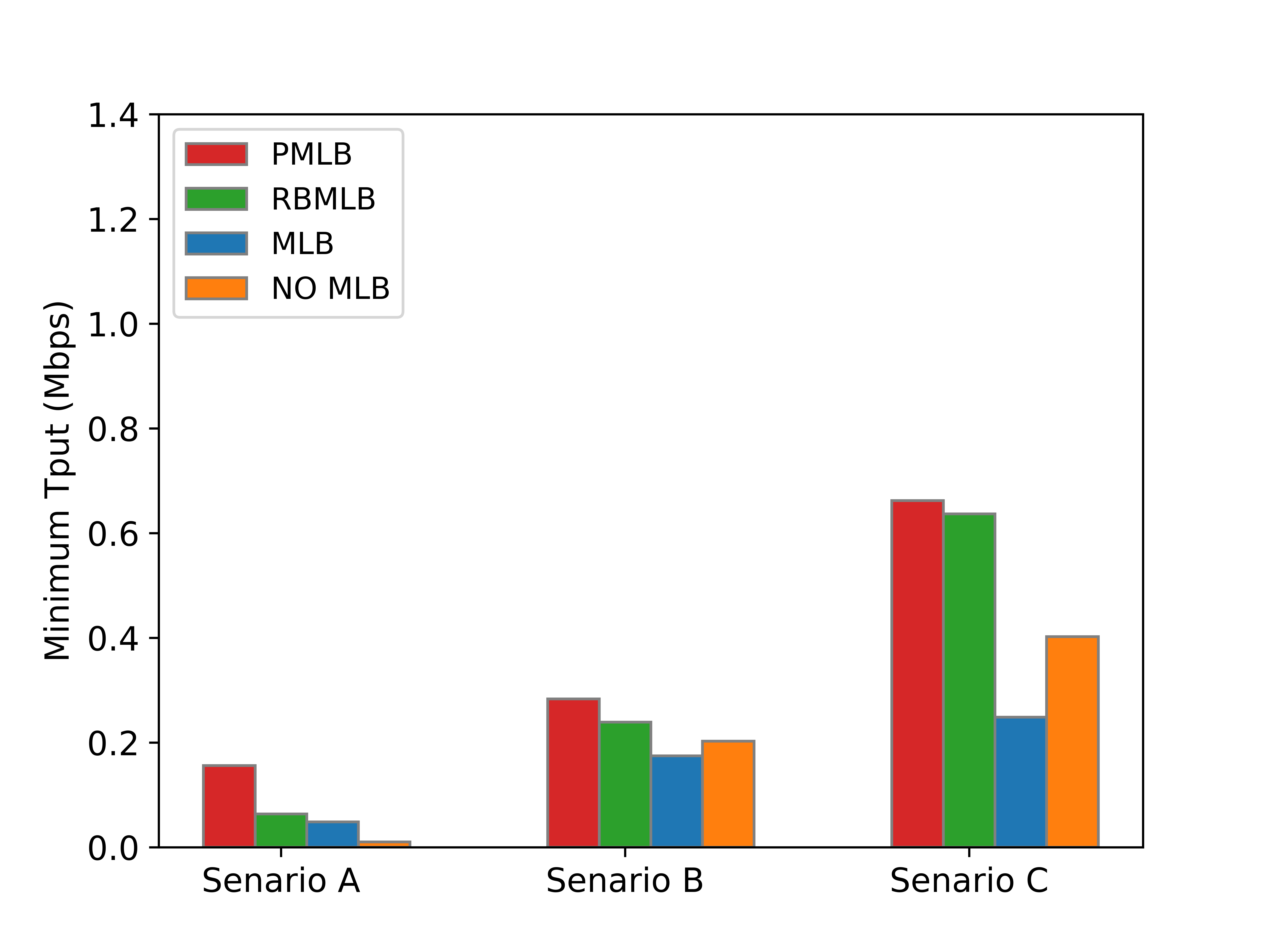}}
\subfigure[\hspace{-.3in}]{\label{fig:4}\includegraphics[scale=0.39,trim={0.2in 0 0.6in 0.54in},clip]{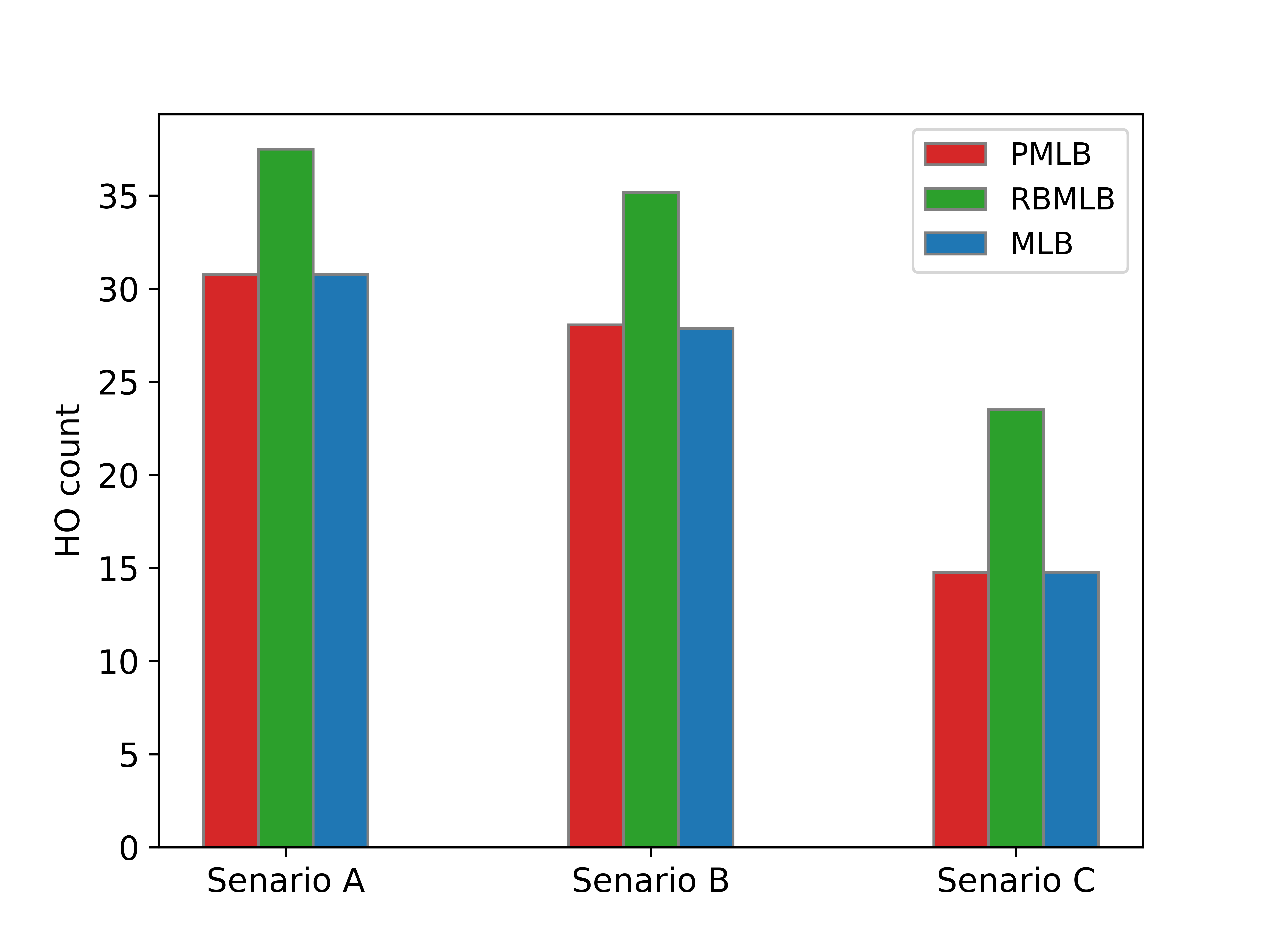}}
\subfigure[\hspace{-.3in}]{\label{fig:5}\includegraphics[scale=0.39,trim={0.2in 0 0.6in 0.54in},clip]{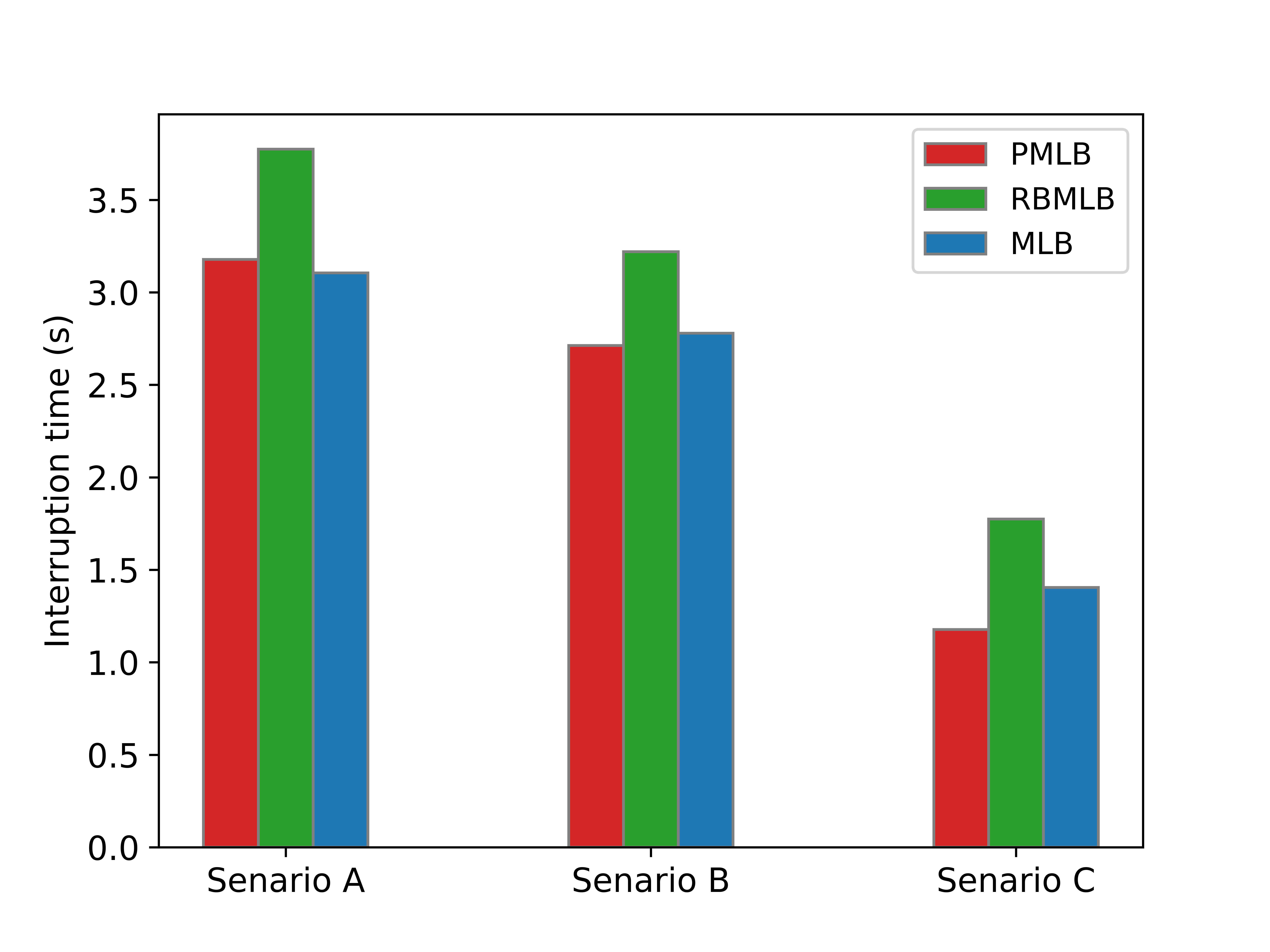}}
\subfigure[\hspace{-.3in}]{\label{fig:6}\includegraphics[scale=0.39,trim={0.2in 0 0.6in 0.54in},clip]{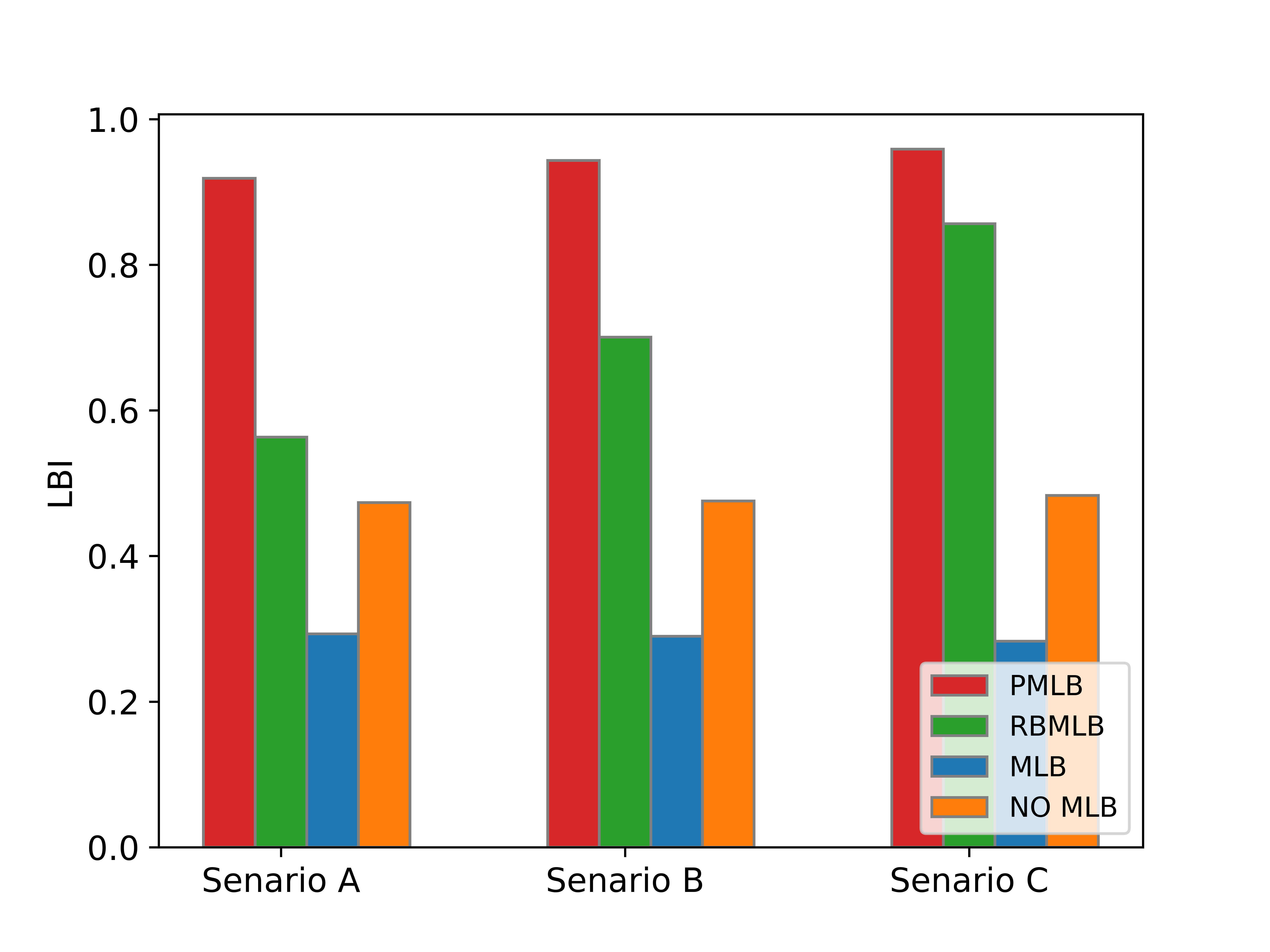}}
\caption{Performance comparison between algorithms in terms of (a) load balancing behavior (b) average throughput (c) minimum throughput (d) HO count (e) HO interruption time (f) LBI}
\end{figure*}

First, we show the algorithm behavior in terms of band utilization and load balancing in Fig. 3(a). It can be noticed that our proposed algorithm utilizes all the bands and distributes the load among the bands the most, followed by RBMLB, whereas the vanilla MLB algorithm utilized mainly two bands which have the best channel quality. Afterwards, the average and minimum cell throughput across the bands are depicted in Fig. 3(b) and Fig. 3(c), respectively. In heavy traffic scenarios, we can see that PMLB algorithm outperforms  RBMLB, achieving an 18\% improvement in the average throughput, and almost double the performance in terms of minimum throughput. This is due to the fact that the PMLB algorithm relocates UEs to bands with best resources or best channel quality in order to guarantee a minimum data rate requirement, unlike the other algorithms. Similarly, in moderate and light traffic scenarios,  PMLB algorithm outperforms other algorithms, but the gap between PMLB and RBMLB almost diminishes in light traffic scenario. In fact, since  PMLB tries to utilize all the bands and distribute the load evenly, some bands might have more capacity to accommodate more UEs without affecting the assigned UEs, and hence, utilizing all the bands equally might not be necessary in light traffic scenarios. Subsequently, the inter-frequency HO count and the cause interruption time are shown in Fig. 3(d) and Fig. 3(e), respectively. The PMLB algorithm achieves less HO count and interruption time than the RBMLB method by 14\%-36\% in all scenarios. This is due to the LBI event that was added as a trigger to the optimization in order to avoid relocating UEs unnecessarily. We can see the LBI values in Fig. 3(f). We set the LBI threshold to $L_{th}$ to 0.8, and it can be seen that the PMLB method was able to maintain an LBI average above 0.9 in all scenarios.

\section{Conclusion}\label{conclusion}
\addtolength{\topmargin}{0.19in}
We have presented a new approach for MLB in multi-band networks, namely, Probabilistic Mobility Load Balancing (PMLB). We have modeled the UE-band assignment problem as an integer multi-objective stochastic optimization problem, where the objectives are to minimize the maximum load and the number of handovers required. To reduce the solution complexity, we have proposed to estimate the random variables and relax the integrality of the problem, and then we have transformed the problem into an LP. The solution outputs a probability distribution for each UE over the bands, where the assigned band to a UE is sampled from that distribution. The simulation results show that the proposed algorithm outperforms rule-based methods in terms of throughput and the interruption time caused by the handovers.


{
\bibliographystyle{IEEEtran}
\bibliography{ref}
}
\end{document}